# System integration and control design of a maglev platform for space vibration isolation

**Zhaopei Gong, Liang Ding, Honghao Yue, Haibo Gao, Rongqiang Liu, Zongquan Deng and Yifan Lu**

## Abstract

Micro-vibration has been a dominant factor impairing the performance of scientific experiments which are expected to be deployed in a micro-gravity environment such as a space laboratory. The micro-vibration has serious impact on the scientific experiments requiring a quasi-static environment. Therefore, we proposed a maglev vibration isolation platform operating in six degrees of freedom (DOF) to fulfill the environmental requirements. In view of the noncontact and large stroke requirement for micro-vibration isolation, an optimization method was utilized to design the actuator. Mathematical models of the actuator's remarkable nonlinearity were established, so that its output can be compensated according to a floater's varying position and a system's performance may be satisfied. Furthermore, aiming to adapt to an energy-limited environment such as space laboratory, an optimum allocation scheme was put forward, considering that the actuator's nonlinearity, accuracy, and minimum energy-consumption can be obtained simultaneously. In view of operating in 6-DOF, methods for nonlinear compensation and system decoupling were discussed, and the necessary controller was also presented. Simulation and experiments validate the system's performance. With a movement range of $10 \times 10 \times 8$ mm and rotations of 200 mrad, the decay ratio of −40 dB/Dec between 1 and 10 Hz was obtained under close-loop control.

## 1. Introduction

Micro-vibration has been a dominant factor impairing space scientific experiments' performance. It contaminates the microgravity environment and degrades the ideal working condition of sensitive payloads (Stark and Stavrinidis, 1994). The term micro-vibration refers to a low-level acceleration disturbance in the microgravity environment, with a wide frequency range (from direct current (d.c.) to 1 kHz) and small magnitude (1 μg to 100 mg, where $g = 9.8\,\mathrm{ms}^2$) (Aglietti et al., 2004). Low frequency disturbance sources include solar panel flutter, gravity gradient drag, and imbalance of flywheels. High frequency sources include crew activity and air conditioning, etc. They are sufficient to degrade the performance of ultra-precision payloads such as a high-resolution camera or long-range laser communication (Li et al., 2018).

Vibration isolation schemes have been developed over recent decades (Liu et al., 2015). While high frequency disturbance can be suppressed utilizing passive methods, active control is a good candidate for low frequency micro-vibration isolation. It can achieve superior performance in low frequency and adapt to various environments (Hansen et al., 2012; Stabile et al., 2017a). Active techniques have been studied and developed to counteract external disturbance (Spanos et al., 1995; Kamesh and Pandiyan, 2010; Wang et al., 2018).

Several kinds of actuators have been applied to achieve active control. Due to high accuracy and fast response, a piezoelectric actuator is widely used in

State Key Laboratory of Robotics and System, Harbin Institute of Technology, People's Republic of China

**Corresponding author:**
Liang Ding, School of Mechatronics Engineering, Harbin Institute of Technology, Harbin, Heilongjiang Province 150001, People's Republic of China Email: liangding@hit.edu.cn

positioning and vibration control (Li and Xu, 2010; Huang and Xu, 2014). However, it can only provide displacement of a few micrometers and is not suitable for micro-vibration isolation, which requires a stroke of millimeters. Besides, features such as creep and sensitivity to environment also limit its application. Parallel kinematic structures and active–passive hybrid structures were also widely investigated (Hauge and Campbell, 2004; Preumont et al., 2007; Wu et al., 2018). Since they are contact type actuators transmitting vibration themselves, their contribution for micro-vibration suppression is limited. A novel negative-resistance electromagnetic shunt damper based on 2-collinear-DOF strut was developed. It showed a superior performance in vibration isolation (Stabile et al., 2017b).

In view of noncontact and large stroke, a maglev actuator seems to be a preferable solution for space micro-vibration isolation. The stator's vibration cannot be transmitted to the floater since there is no stiction or friction between the moving and stationary parts. Previous research shows that maglev actuators are capable of achieving noncontact, high resolution, and large stroke. Verma et al. (2004) presented a 6-DOF positioner which has 5 nm resolution and 1 Kg load capacity over a travel range of 300 μm. The design is worth considering, but it is designed for special application, and stroke is relatively small. Kim and Verma (2007) designed a positioner with novel electromagnetic actuator, which has 4 nm resolution and 5 mm travel range. It is more suitable to be applied for small size and light mass missions. Mei-Yung Chen and colleagues developed a 6-DOF maglev positioner with fluid bearing, resulting in 4 mm stroke (Chen et al., 2011). However, the system's construction is complex and the fluid part is not feasible in space. A maglev lens driving actuator which has 10 mm stroke with tracking error of less than 12 μm was developed by Dongjue He and colleagues (He et al., 2013). The design concept can enhance magnetic flux density but the structure is complex, and the total weight will be increased.

In view of noncontact and large stroke, we presented a design of maglev actuator. Based on it, a maglev vibration isolation platform (MVIP) was put forward. Efforts were focused on three aspects: (1) establishing the actuator's nonlinearity mathematical model; (2) minimizing system power consumption; and (3) compensating the system nonlinearity and decoupling system. Considering the actuator's position-dependent characteristic, a static experiment and fitting process were carried out to obtain the mathematical model. Aiming to adapt to an energy-limited environment, such as a space laboratory operating in orbit, optimization theory was adopted to allocate actuation efforts. Besides, an online recursive least squares (RLS) method was analyzed and utilized to rectify the system's cross-coupling.

This paper consists of six sections. In Section 2, design concepts of actuator as well as MVIP are presented. In Section 3, the actuator's mathematical model and the dynamics of whole system are analyzed and formulated. In Section 4, the proper design of nonlinearity compensation, optimum actuation allocation, and controller are described. An online rectification method for dynamics coupling was also discussed. To verify the system's performance, experiments are carried out in Section 5. Finally, conclusions are summarized in Section 6.

## 2. Configuration of maglev platform

A vibration isolation platform utilizing Lorentz forces was designed and made to achieve micro-vibration isolation in 6-DOF. In order to improve the actuator's performance, an optimization method for structural design was adopted. A measurement system was also developed to acquire the platform's movement and feedback this to the controller.

### 2.1. Platform design and actuation scheme

To satisfy the requirement of micro-vibration isolation, a platform-level system was put forward as shown in Figure 1. In general, MVIP mainly consists of a floater, a stator, eight actuators, a measurement system, and a controller. The stator is equipped with three position sensitive detector (PSD) sensors, three mutually perpendicular accelerometers, and eight magnet yokes. The floater is equipped with three accelerometers and eight coils. The floater has no mechanical contact with the stator besides the umbilical cable, so it can move freely within the stator's frame. Without loss of generality, the cable is modeled as spring-damper, which will be discussed later. Thus, this scheme avoids vibration transmission caused by mechanical contact.

Eight actuators distribute on four sides of the MVIP as illustrated in Figure 2. Four actuators provide force along the $z$-axis and torques around the $x$, $y$-axes. Other actuators provide force along $x$, $y$-axes as well as torque around the $z$-axis. This configuration enables the floater's ability of operating in 6-DOF. Actuators' output force, which are denoted as $\boldsymbol{F}_{motor}$, constitute the actuation effort on the floater's center of mass (CoM).

We defined the forces to be $f_1$, $f_3$, $f_5$, and $f_7$ as the forces generated by horizontal actuators, and $f_2$, $f_4$, $f_6$, and $f_8$ generated by vertical actuators. The resultant forces, denoted as $\boldsymbol{F}_C = \begin{bmatrix} F_{Cx} & F_{Cy} & F_{Cz} \end{bmatrix}^T$, and resultant torques, denoted as $\boldsymbol{\tau}_C = \begin{bmatrix} \tau_{Cx} & \tau_{Cy} & \tau_{Cz} \end{bmatrix}^T$, can be derived as follows

$$\begin{bmatrix} \boldsymbol{F}_C \\ \boldsymbol{\tau}_C \end{bmatrix} = \underset{6\times 8}{\boldsymbol{C}_K} \boldsymbol{F}_{motor} \tag{1a}$$

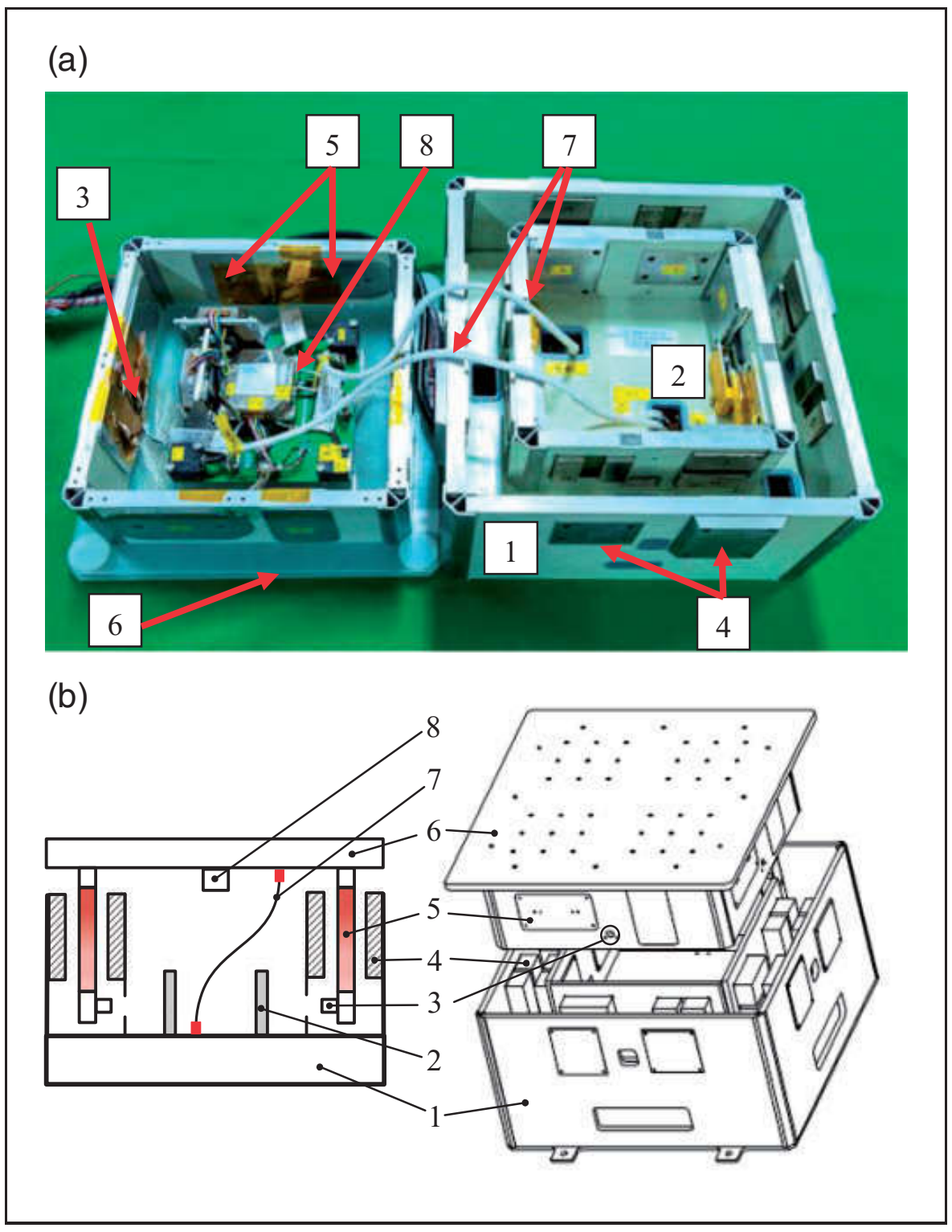

**Figure 1.** Exploded view of maglev vibration isolation platform: (1) stator; (2) position sensitive detector sensor; (3) laser source; (4) magnets; (5) coils; (6) floater; (7) cable; and (8) accelerometer.

$$\begin{bmatrix} F_{Cx} \\ F_{Cy} \\ F_{Cz} \\ \tau_{Cx} \\ \tau_{Cy} \\ \tau_{Cz} \end{bmatrix} = \begin{bmatrix} 1 & 0 & 0 & 0 & -1 & 0 & 0 & 0 \\ 0 & 0 & 1 & 0 & 0 & 0 & -1 & 0 \\ 0 & 1 & 0 & 1 & 0 & 1 & 0 & 1 \\ 0 & L_1 & 0 & -L_2 & 0 & -L_1 & 0 & L_2 \\ 0 & -L_2 & 0 & -L_1 & 0 & L_2 & 0 & L_1 \\ -L_1 & 0 & L_1 & 0 & -L_1 & 0 & L_1 & 0 \end{bmatrix} \times \begin{bmatrix} f_1 \\ f_2 \\ f_3 \\ f_4 \\ f_5 \\ f_6 \\ f_7 \\ f_8 \end{bmatrix} \quad (1b)$$

where $L_1$ and $L_2$ are the moment arms, respectively.

For a space laboratory facility, sufficient redundancy should be provided. Thus, we adopt the over-actuation scheme here in case of electronic components failure caused by a high energy cosmic ray. Besides, this scheme lowers the maximum output force of single actuator and increases the system's robustness. In different failure modes, as long as the matrix in equation (1) is row full rank, the system's function would be maintained. In one-actuator failure mode, effects on the system are limited. In two-actuator failure mode, except the simultaneous failure of No.1/No.5 or No.3/No.7 pair, the system's performance can be maintained by increasing the other actuators' current. However, suffering high amplitude or impulse vibration, performance may be degraded. Until more than two actuators fail, the system will turn into a security mode, locking the floater with stator tightly.

## 2.2. Maglev actuator design

A maglev actuator was put forward to fulfill the requirements of noncontact and large stroke. As depicted in Figure 3, the actuator consists of a square coil, nesting inside the floater's flank, and two pairs of permanent magnets, mounted on the stator. In this scheme, force can be adjusted by current flowing in the coil.

Since there should be no contact between moving and stationary parts, stroke is mainly determined by the ambient vibration acceleration $a_{vib}$ and frequency $f_{vib}$, as

$$\boldsymbol{x}_{rms} = \frac{a_{vib}}{(2\pi f_{vib})^2} \quad (2)$$

According to the space missions' record (DeLombard and Hakimzadeh, 1997), environment disturbance reaches its maximum amplitude of 0.1 g with a frequency of 3 Hz. The stroke evaluates to $\boldsymbol{x}_{rms} = 2.8\,\text{mm}$ and the peak to peak value is $\boldsymbol{x}_{peak-peak} = 8\,\text{mm}$. Leaving some margin, the stroke should be larger than 10 mm. This requirement is beyond most existing actuators' ability, as mentioned in Section 1. We put forward a design of actuator according with previous work (Wu et al., 2014). The actuator's basic requirements are given in Table 1.

Aiming to make this platform adapt to a space environment, a parametric study was carried out to fulfill three conditions simultaneously: (1) minimizing heat consumption; (2) minimizing coil's weight; and (3) maximizing magnetic flux. The actuator structural parameters are presented in Table 2. This study can be attributed to a multi-objective optimization design expressed as

$$Opt.\begin{cases} Max : B \\ Min : Q \\ Min : m_{coil} \end{cases} \quad (3a)$$

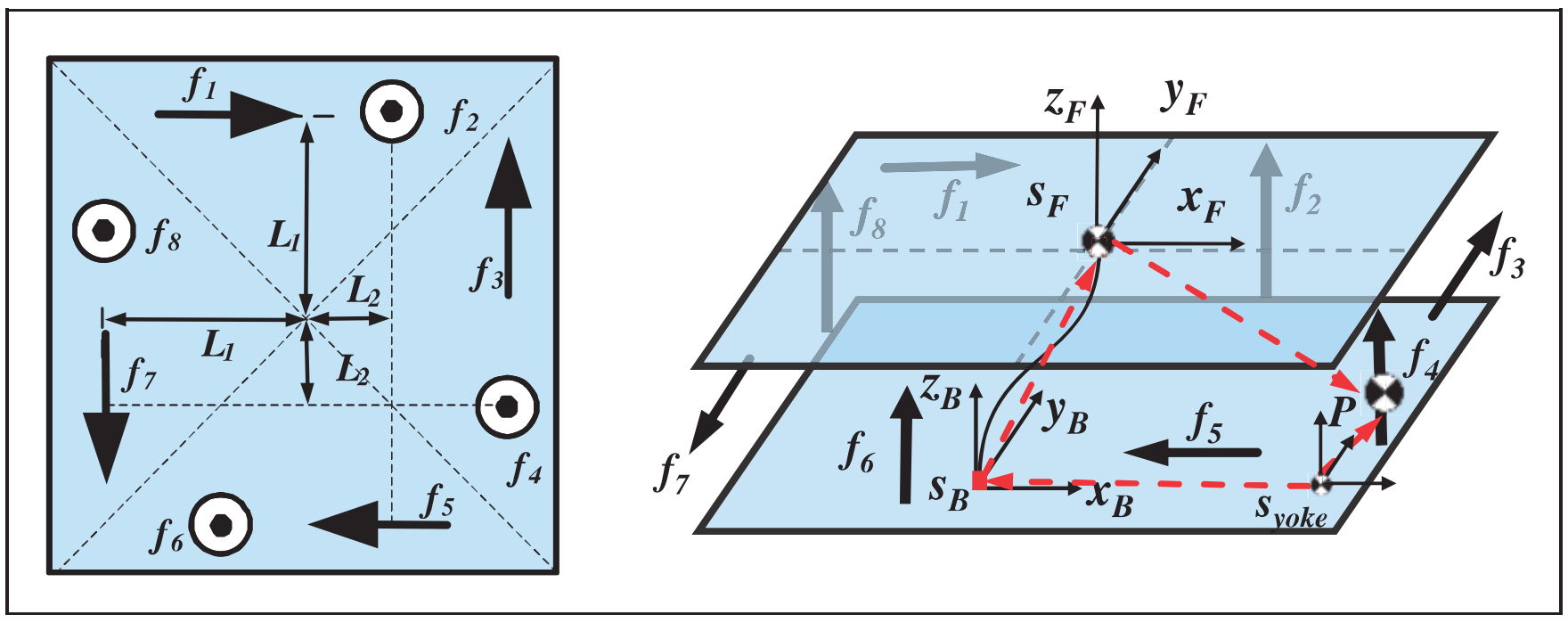

**Figure 2.** Platform coordinate and actuation scheme.

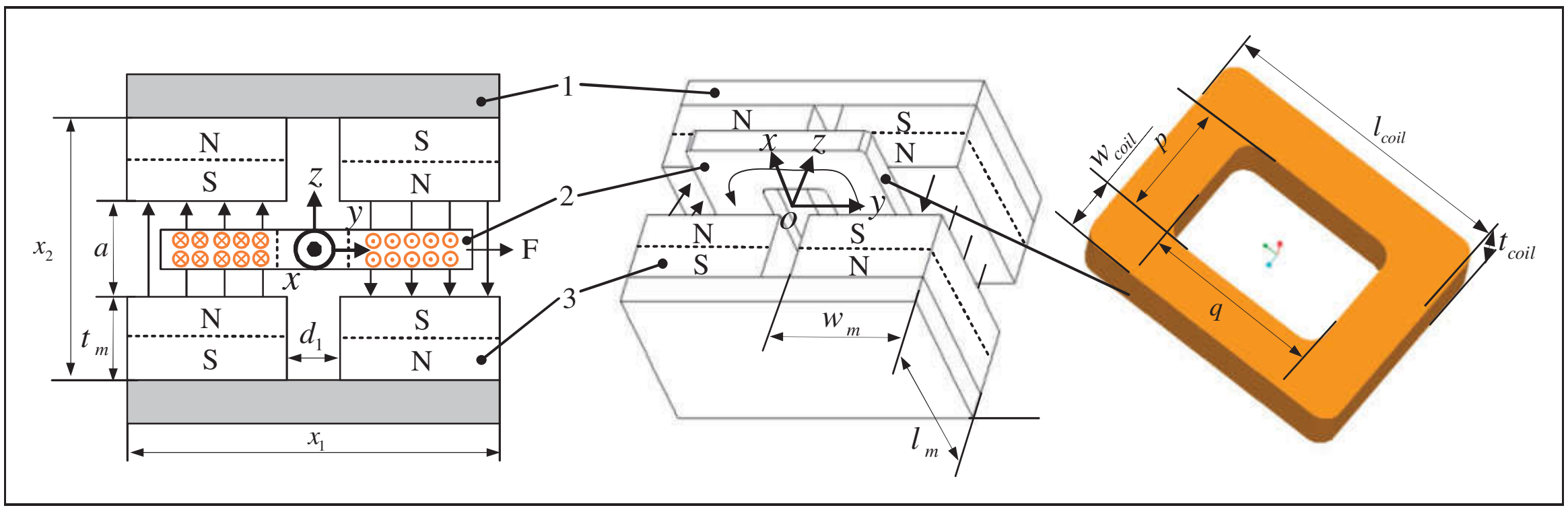

**Figure 3.** 1-degree of freedom maglev actuator: (1) magnet yoke; (2) coil; and (3) permanent magnet.

**Table 1.** Actuator's basic requirements.

| Design index | Parameter | Value |
|---|---|---|
| Maximum output force | $F_{max}$ | $\geq 24\,N$ |
| Maximum current | $I_{max}$ | 2 A |
| Minimum stroke | $s$ | $\geq 10\,mm$ |

$$s.t.\begin{cases} 2t_m + a - x_2 \leq 0 \\ 2w_m + d_1 - x_1 \leq 0 \\ l_{coil} - x_3 \leq 0 \\ F_{max} = 2N_{coil}BI_{max}l_m \geq 24 \end{cases} \tag{3b}$$

where magnetic flux ($B$) is an empirical constant expressed in equation (4) and $B_r$ denotes the remanence. $t_m$, $w_m$, $d_1$ and $l_{coil}$ are dimension parameters as illustrated in Figure 3. $x_1$, $x_2$ and $x_3$ are the maximum size of the actuator's width, thickness, and height, corresponding to 65, 55, and 70 mm. The purpose of maximizing $B$ is that we want to obtain a relatively large output force subject to a relatively small current, and a relatively short coil length. A group of boundary values for the magnet's size were set, so the value of magnetic flux ($B$) will not keep growing.

$$B = \frac{2B_r}{\pi}\left[\tan^{-1}\left(\frac{w_m l_m}{a\sqrt{a^2 + l_m^2 + w_m^2}}\right) - \tan^{-1}\left(\frac{w_m l_m}{(4t_m + a)\sqrt{(4t_m + a)^2 + l_m^2 + w_m^2}}\right)\right] \tag{4}$$

The coil's heat dissipation $Q$ and mass $m_{coil}$ are derived in equation (5) and equation (6).

$$Q = \frac{16 I_{max}^2 \rho_r V_{coil}}{\pi^2 d_{coil}^4} \tag{5}$$

$$m_{coil} = \rho \eta_{pack} V_{coil} \tag{6}$$

where

$$V_{coil} = \frac{\pi}{2}\cdot d_{coil}^2\left(\frac{a-s-c}{d_{coil}}\right)\left(\frac{N_{coil}d_{coil}}{a-s-c}(p+q+4d_{coil}) + \frac{N_{coil}d_{coil}^2}{a-s-c}\left(\frac{N_{coil}d_{coil}}{a-s-c} - 1\right)\right) \tag{7}$$

**Table 2.** Actuator structure parameters.

| Design index | Parameter | Value | Design index | Parameter | Value |
|---|---|---|---|---|---|
| Magnet's length (mm) | $l_m$ | 40.5 | Coil's width (mm) | $w_{coil}$ | 11 |
| Magnet's width (mm) | $w_m$ | 20.5 | Coil's length (mm) | $l_{coil}$ | 64 |
| Magnet's thickness (mm) | $t_m$ | 14 | Coil's thickness (mm) | $t_{coil}$ | 12 |
| Air gap (mm) | $a$ | 26 | Stroke (mm) | $s$ | 12 |
| Heat dissipation (W) | $Q$ | 6.85 | Coil's mass (g) | $m_{coil}$ | 443 |
| Magnetic flux (T) | $B$ | 0.4 | Turns of coil | $N_{coil}$ | 380 |

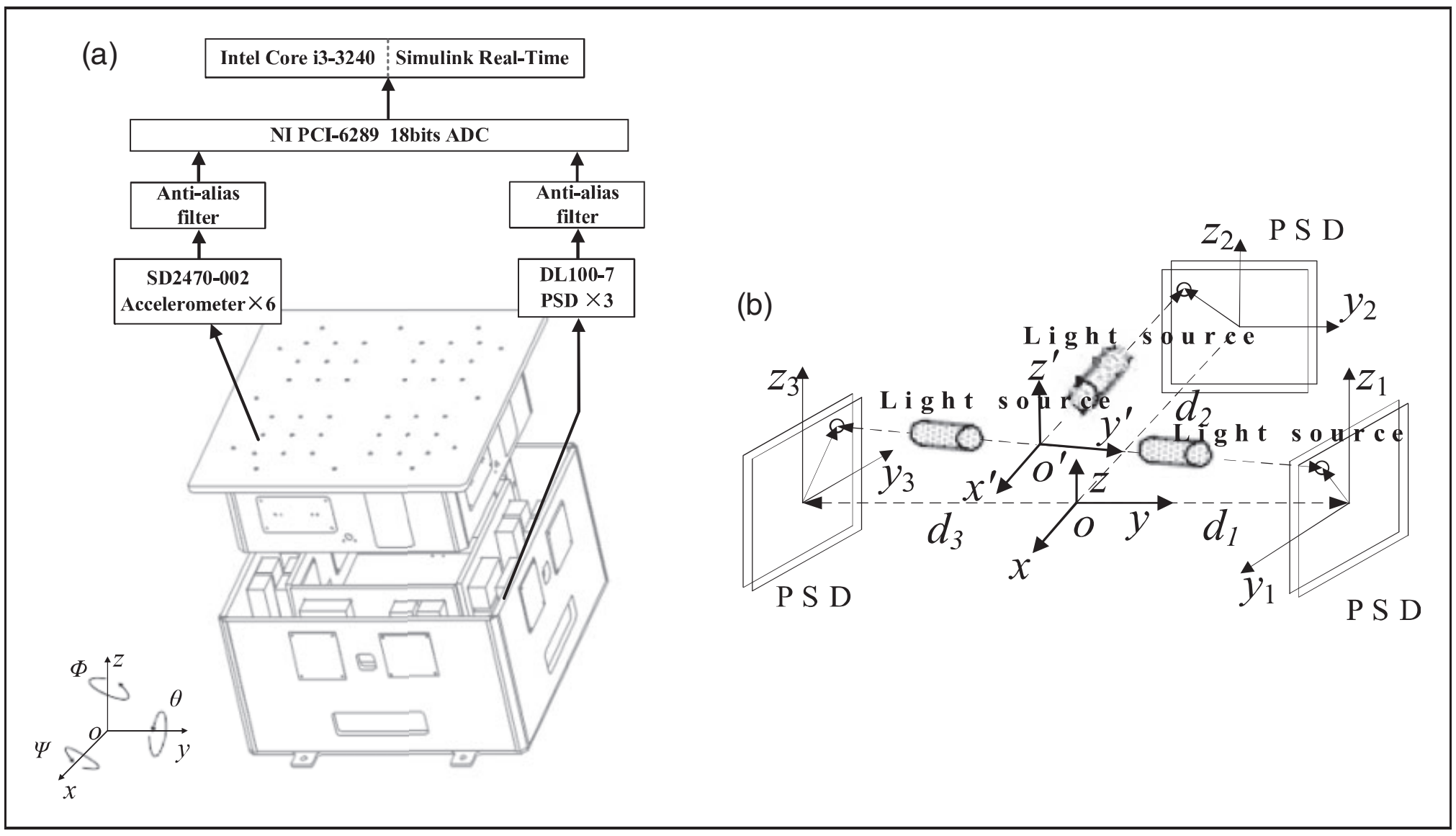

**Figure 4.** (a) Measurement system integration; and (b) arrangement of position sensitive detector (PSD) sensors.

$d_{coil}$ denotes the copper wire's diameter, $\rho_r$ is resistivity of copper, $c$ denotes the coil supporting frame's thickness, $\rho$ denotes the density of copper which is a constant, and $\eta_{\text{pack}}$ the filing ratio of copper alloy. Since a genetic algorithm can find a globally optimal resolution efficiently, the design parameters ($w_{coil}$, $l_{coil}$, $N_{coil}$, $l_m$, $w_m$, $t_m$, $s$) were solved out by utilizing the MATLAB optimization toolbox and are presented in Table 2.

## 2.3. Sensor instrumentation integration

The floater needs to remain stable within the stator's frame and attenuate vibration simultaneously. Thus, a sensing system was established for providing motion to the controller. As shown in Figure 4 (a), a sensing system, consisting of three PSD sensors and six accelerometers, was developed and integrated with the MVIP. PSD sensors are mounted on the stator, while corresponding light sources are mounted on the floater. The floater's displacement is measured by PSD sensors as shown in Figure 4 (b).

According to the coordinate in Figure 4 (a), the relationship between sensors' output ($p_{1y} - p_{3z}$) and floater's displacement can be expressed as

$$\begin{bmatrix} p_{1y} \\ p_{2y} \\ p_{3y} \\ p_{1z} \\ p_{2z} \\ p_{3z} \end{bmatrix} = \left[\begin{array}{ccc|ccc} 1 & 0 & -d_1 & 0 & 0 & 0 \\ 0 & 1 & -d_2 & 0 & 0 & 0 \\ -1 & 0 & -d_3 & 0 & 0 & 0 \\ \hline 0 & 0 & 0 & 1 & d_1 & 0 \\ 0 & 0 & 0 & 1 & 0 & d_2 \\ 0 & 0 & 0 & 1 & -d_3 & 0 \end{array}\right] \begin{bmatrix} \Delta x \\ \Delta y \\ \Delta \phi \\ \Delta z \\ \Delta \psi \\ \Delta \theta \end{bmatrix} \tag{8}$$

where $d_1$, $d_2$, and $d_3$ are distances from each sensor to the floater's geometrical center, therefore, the floater's relative displacement can be calculated from

$$\begin{bmatrix} \Delta x \\ \Delta y \\ \Delta \phi \\ \Delta z \\ \Delta \psi \\ \Delta \theta \end{bmatrix} = \left[\begin{array}{ccc|ccc} d_3/(d_1+d_3) & 0 & -d_1/(d_1+d_3) & 0 & 0 & 0 \\ -d_2/(d_1+d_3) & 1 & -d_2/(d_1+d_3) & 0 & 0 & 0 \\ -1/(d_1+d_3) & 0 & -1/(d_1+d_3) & 0 & 0 & 0 \\ \hline 0 & 0 & 0 & d_3/(d_1+d_3) & 0 & d_1/(d_1+d_3) \\ 0 & 0 & 0 & 1/(d_1+d_3) & 0 & -1/(d_1+d_3) \\ 0 & 0 & 0 & -d_3/(d_2\cdot(d_1+d_3)) & 1/d_2 & -d_1/(d_2\cdot(d_1+d_3)) \end{array}\right] \times \left[p_{1y} \quad p_{2y} \quad p_{3y} \quad p_{1z} \quad p_{2z} \quad p_{3z}\right]^T \quad (9)$$

By properly arranging sensors, a 100 nm translation and 0.01 mrad rotation sensitivity can be achieved. Acceleration of the stator as well as floater can be obtained by adopting proper installation of accelerometers and a solution algorithm. More details about these schemes can be found in our previous work (Wu et al., 2015; Zhu et al., 2015).

To satisfy the MVIP's performance, Simulink Real-Time was adopted and implemented on a Core i3 processor as a real-time controller. Sensor signals are digitized with an 18-bit analog-to-digital converter, while actuators are commanded with a 16-bit digital-to-analog converter. A graphic interface runs on a personal computer, while connecting to the real-time controller via Transmission Control Protocol/Internet Protocol. The controller runs with a close-loop frequency of 2 kHz.

## 3. Mechatronic system modeling

System performance is limited by different nonlinear terms caused by the actuator, cables and dynamics. Since a low-jitter environment is desired, the modeling of nonlinearity and overall effect acting on the floater's dynamics, is indispensable.

### 3.1. Mathematical model of nonlinear maglev force

According to the Lorentz equation, the actuation force between current-flowed coil and a ferromagnetic is calculated by

$$f = \int (\boldsymbol{J} \times \boldsymbol{B}) \quad (10)$$

where $\boldsymbol{J}$ is current density, $\boldsymbol{B}$ is magnetic flux density, and $f$ is Lorentz force. Under a general situation, the magnetic flux is not homogenous in space and such a description of current-to-force relation includes a quintuple integral described in Gu et al. (2004). To realize precise output within whole stroke, it is essential to adopt a method, which is feasible and easy calculating, to figure out the current based on desired output force and the coil's position. Due to the difficulty in deriving an easy calculating expression from the quintuple integral, a finite element modeling (FEM) simulation and calibration experiment were carried out to obtain and verify the expression. The FEM simulation results show that the coil's rotation ($< 5°$) can be neglected since its effect is limited.

As shown in Figure 5, the calibration unit consists of a current source, a maglev actuator, two force sensors, two data acquisition units, and a 3-DOF positioning platform. The current in coil was fixed at 1.0 A. Force sensors are mounted on the coil and the yoke were adjust by 3-DOF platform. Among whole work space ($10 \times 10 \times 10\,mm^3$), force was measured by each 1mm, so there are 1000 points of data in all.

As shown in Figure 6 (a), output forces according position variation along the $y$ and $z$-axis are obtained from FEM and experiment, respectively, and the measured data are around 92% of the simulated. For $x$-axis, since the coil's length is longer than stroke, position variation has no effect on output force. Along $z$-axis, the output force reaches its minimum at center, while the opposite occurs along $y$-axis. The heterogeneity of magnetic field is obvious, and the output varies up to 12%.

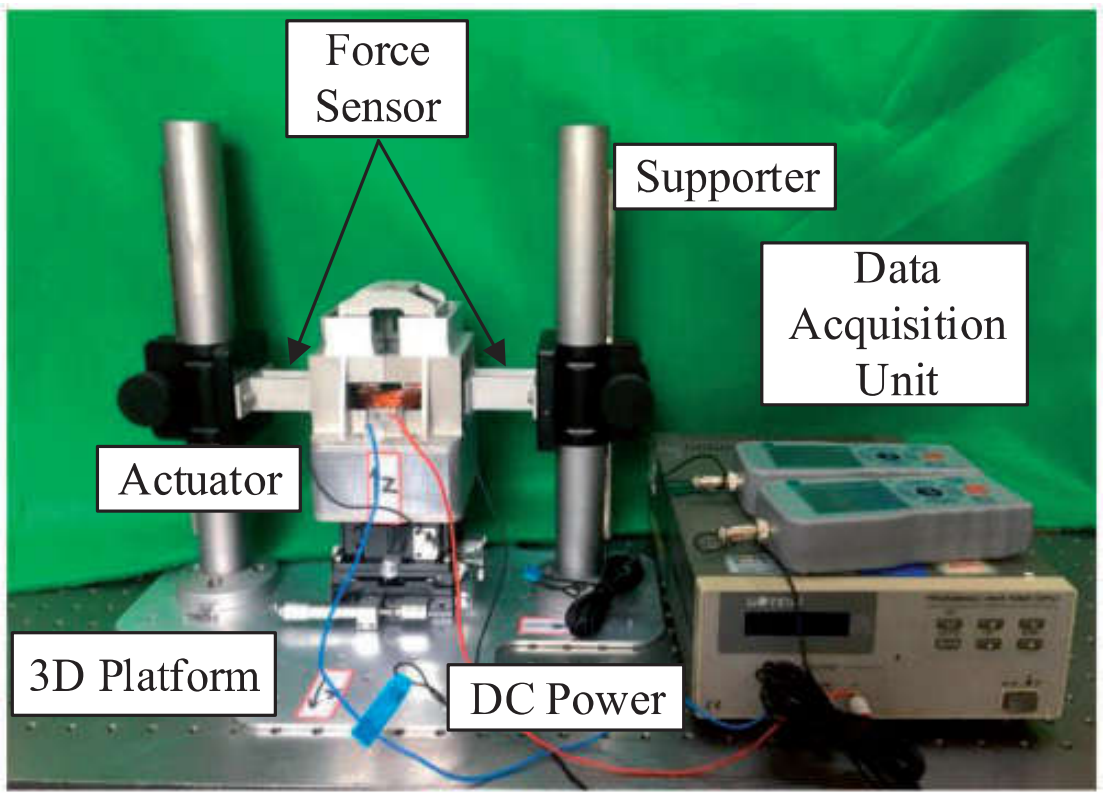

**Figure 5.** Static calibration experiment unit.

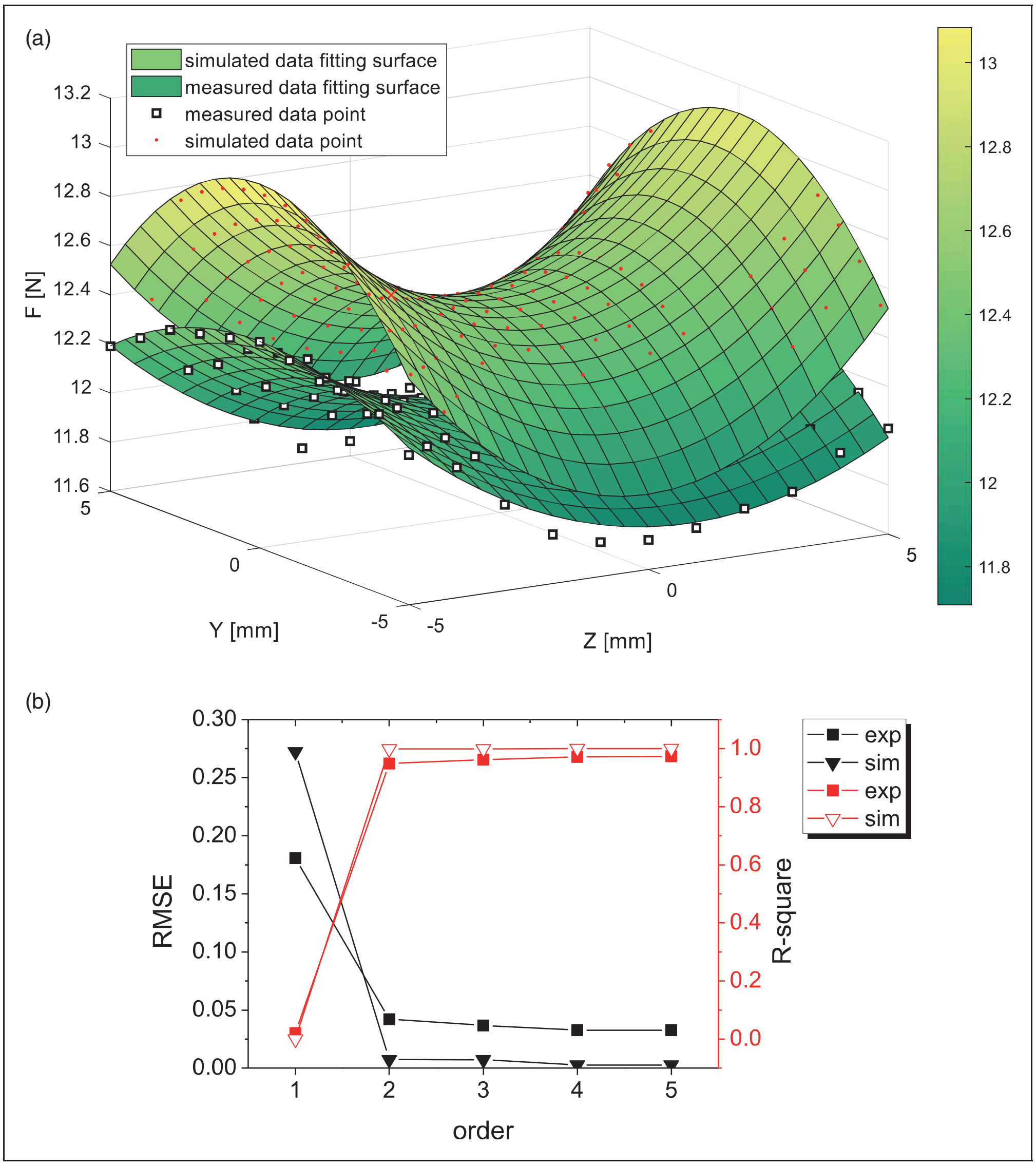

**Figure 6.** Establishment of actuator's mathematical model: (a) actuation force as a function of space variables obtained by finite element modeling and experiment; and (b) fitting results under different orders.

Since the magnetic field is heterogeneity, the current-to-force relationship may be written as $F_i/I_i = Q_i(y_i, z_i)$ (i = 1to8), where $F_i$, $I_i$ and $Q_i$ denotes the $i$-th actuator's output force, current and model. $y_i$, $z_i$ denote coil's position along $y$, $z$-axes.

According to variation in the data, a hyperbolic paraboloid can be utilized to present the actuator's model $Q_i(y_i, z_i)$. The data were fitted by utilizing the polynomial fitting function in MATLAB and the results are shown in Figure 6 (a). For obtaining a balance between precision and calculating, different orders were set and results are summarized in Figure 6 (b). For the experimental data, due to the measurement error, R-square increase and root-mean-square error decrease with the growth of order which means the fitting error declines. However, 3-order was chosen here to avoid overfitting. The fitting function $Q_i$ is described as follows

$$Q_i(y_i, z_i) = [1, y_i, z_i, y_i^2, y_i z_i, z_i^2, y_i^3, y_i^2 z_i, y_i z_i^2, z_i^3] \cdot \boldsymbol{G}^T \tag{11}$$

where $\boldsymbol{G} \in \boldsymbol{R}^{1\times10}$ is the coefficient matrix obtained by MATLAB; thus, the input current can be calculated in real-time according to the coil's position and desired output force.

### 3.2. Dynamics of MVIP

Dynamics of MVIP can be derived using the Newton–Euler equation following three assumptions: (1) the floater is regarded as rigid body; (2) floater's CoM is coincident with its geometric center due to symmetrical design; and (3) cable is attached on the floater's CoM.

In Figure 2, coordinate $S_F$ is attached to the floater's CoM while another coordinate $S_B$ is attached to the stator. Rotation matrix ${}^B_F\boldsymbol{R}$ and translation matrix ${}^B\boldsymbol{P}_{FORG}$ are terms of transfer matrix ${}^{S_B}_{S_F}T$ to relate these two coordinates

$$\begin{bmatrix} S_B \\ 1 \end{bmatrix} = {}^{S_B}_{S_F}T \cdot \begin{bmatrix} S_F \\ 1 \end{bmatrix} = \left[\begin{array}{c|c} {}^B_F\boldsymbol{R} & {}^B\boldsymbol{P}_{FORG} \\ \hline 000 & 1 \end{array}\right] \begin{bmatrix} S_F \\ 1 \end{bmatrix}$$
$$= \left[\begin{array}{ccc|c} c\theta c\phi & -c\theta s\phi & s\theta & x \\ s\psi s\theta c\phi + c\psi s\phi & -s\psi s\theta s\phi + c\psi c\phi & -s\psi c\theta & y \\ -c\psi s\theta c\phi + s\psi s\phi & c\psi s\theta s\phi + s\psi c\phi & c\psi c\theta & z \\ \hline 0 & 0 & 0 & 1 \end{array}\right] \begin{bmatrix} S_F \\ 1 \end{bmatrix} \tag{12}$$

where $s\phi = \sin\phi$, $c\phi = \cos\phi$, $s\theta = \sin\theta$, $c\theta = \cos\theta$, $s\psi = \sin\psi$, $c\psi = \cos\psi$.

The cable, transmitting power and signal for payloads, is indispensable in the system, and its dynamics can be simplified into spring and damping terms. Thus, disturbance force $\boldsymbol{F}_d$ and torques $\boldsymbol{\tau}_d$ generated by cables are derived as

$$\begin{bmatrix} \boldsymbol{F}_d \\ \boldsymbol{\tau}_d \end{bmatrix} = \underset{6\times6}{\boldsymbol{K}_u} \cdot \begin{bmatrix} {}^B\boldsymbol{P}_{FORG} \\ \boldsymbol{q} \end{bmatrix} + \underset{6\times6}{\boldsymbol{D}_u} \cdot \begin{bmatrix} {}^B\dot{\boldsymbol{P}}_{FORG} \\ \dot{\boldsymbol{q}} \end{bmatrix} \tag{13}$$

where $\boldsymbol{q} = \begin{bmatrix} \alpha & \beta & \gamma \end{bmatrix}^T$ denotes Euler angles. $\boldsymbol{F}_d = \begin{bmatrix} F_{dx} & F_{dy} & F_{dz} \end{bmatrix}^T$, $\boldsymbol{\tau}_d = \begin{bmatrix} \tau_{dx} & \tau_{dy} & \tau_{dz} \end{bmatrix}^T$, $\boldsymbol{K}_u$ and $\boldsymbol{D}_u$ are all diagonal matrixes representing the cable's stiffness and damping.

Based on the Newton–Euler equation, the floater's motion can be derived as

$$\begin{cases} \boldsymbol{M} \cdot {}^B\ddot{\boldsymbol{P}}_{FORG} = \boldsymbol{F}_C + \boldsymbol{F}_d + \boldsymbol{Gr} \\ \boldsymbol{J}_F \cdot \dot{\boldsymbol{\omega}}_F + \boldsymbol{\omega}_F \times (\boldsymbol{J}_F \cdot \boldsymbol{\omega}_F) = \boldsymbol{\tau}_C + \boldsymbol{\tau}_d \end{cases} \tag{14}$$

where $\boldsymbol{M}$ denoting floater's mass is diagonal and $\boldsymbol{J}_F$ denoting floater's inertia is full matrix $3 \times 3$, $\boldsymbol{Gr} = \begin{bmatrix} 0 & 0 & mg \end{bmatrix}^T$ denotes the gravity terms, and $\boldsymbol{\omega}_F = \begin{bmatrix} \omega_{Fx} & \omega_{Fy} & \omega_{Fz} \end{bmatrix}^T$ is floater's instantaneous angular velocity.

Since the relation between Euler angles' velocity $\dot{\boldsymbol{q}}$ and instantaneous angular velocity $\boldsymbol{\omega}_F$ is described as (Greenwood, 2006)

$$\boldsymbol{\omega}_F = \boldsymbol{T}\dot{\boldsymbol{q}} \tag{15a}$$

$$\boldsymbol{T} = \begin{bmatrix} \cos\beta\cos\gamma & \sin\gamma & 0 \\ -\cos\beta\sin\lambda & \cos\gamma & 0 \\ \sin\beta & 0 & 1 \end{bmatrix} \tag{15b}$$

MVIP's dynamics is formulized as

$$\begin{cases} {}^B\ddot{\boldsymbol{P}}_{FORG} = m^{-1}(\boldsymbol{F}_C + \boldsymbol{F}_d + \boldsymbol{G}) \\ \ddot{\boldsymbol{q}} = (J_F\boldsymbol{T})^{-1}[(\boldsymbol{\tau}_c + \boldsymbol{\tau}_d) - \boldsymbol{\omega}_F \times (J_F \cdot \boldsymbol{\omega}_F)] \end{cases} \tag{16}$$

## 4. System decoupling and control strategy

Since MVIP is open loop unstable, the controller is indispensable. Taking MVIP's relative position (between floater and stator) and the floater's acceleration as feedback, the controller maneuvers the floater's motion by adjusting the coil's current. As shown in Figure 7 (a), the control architecture consists of four parts: (1) nonlinear compensation; (2) optimum actuation allocation; (3) dynamics coupling rectification; and (4) controller.

### 4.1. Nonlinear compensation

Nonlinearity of MVIP is caused by actuators, cables, and centrifugal terms. Since the rotation angle is small, nonlinearity caused by rotation can be neglected. For compensating the nonlinearity caused by cable and gravity, an approach is feedback linearization (He and Wei, 2008; Nielsen et al., 2010), which can transform a nonlinear system algebraically into a linear one. This process can be formulated by introducing a virtual

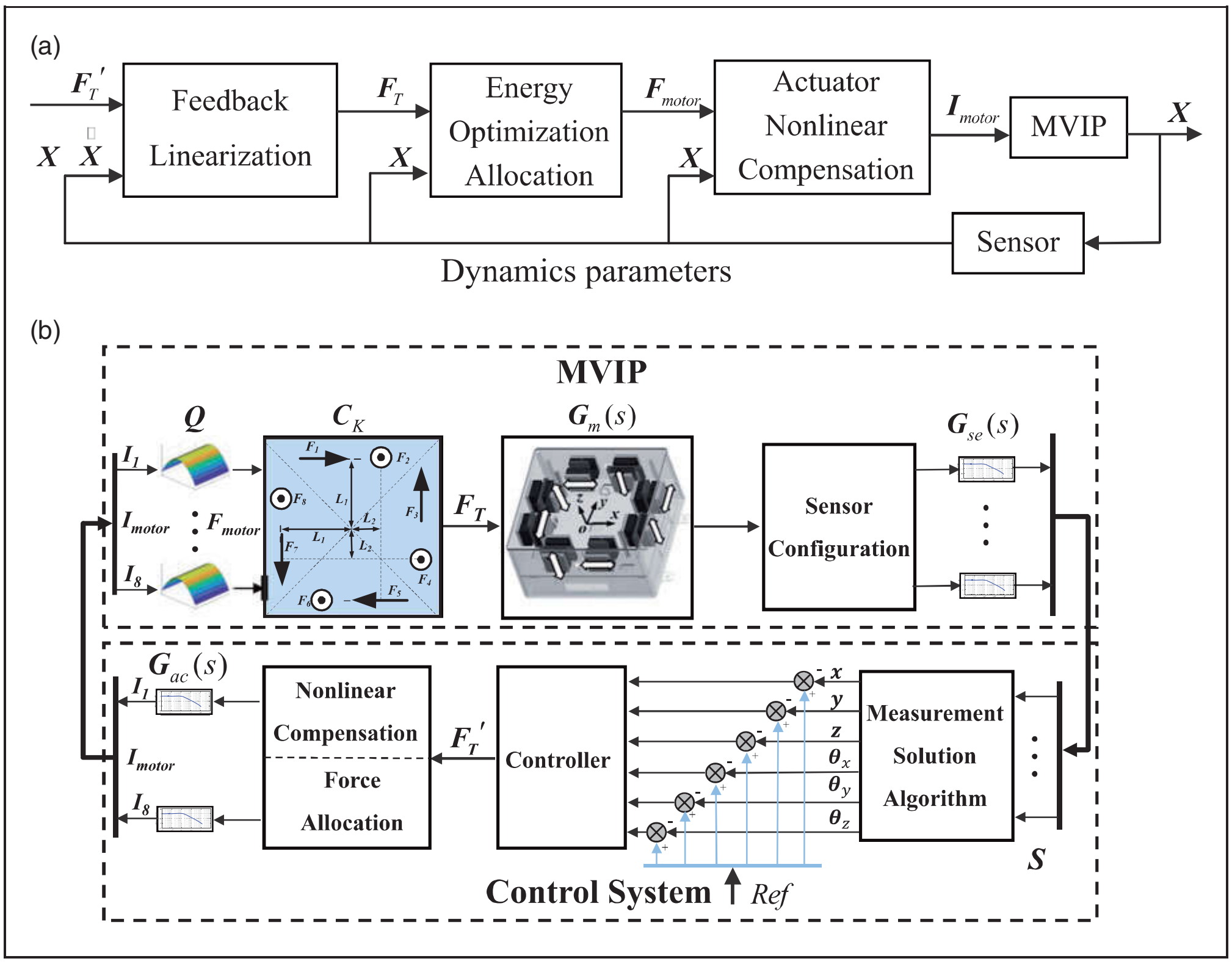

**Figure 7.** (a) Implementation of control strategy; and (b) control architecture of maglev vibration isolation platform (MVIP).

control effort $F_T'$

$$F_T' = F_T - K_u \cdot \begin{bmatrix} {}^{B}P_{FORG} \\ q \end{bmatrix} - D_u \cdot \begin{bmatrix} {}^{B}\dot{P}_{FORG} \\ \dot{q} \end{bmatrix} - \begin{bmatrix} G \\ 0 \end{bmatrix} \tag{17}$$

where $F_T = \begin{bmatrix} F_C^T & \tau_C^T \end{bmatrix}^T$. Then, equation (16) is linearized as

$$\begin{bmatrix} {}^{B}\ddot{P}_{FORG} \\ \ddot{q} \end{bmatrix} = \begin{bmatrix} M & \underset{3\times3}{0} \\ \underset{3\times3}{0} & J_F \end{bmatrix}^{-1} F_T' \tag{18}$$

We utilize the fitting function $Q$ (equation (11)) to compensate for the actuator's nonlinearity. As long as coil's position is obtained, the current can be calculated as

$$I_{motor} = F_{motor} \cdot Q^{-1}({}^{S_{yoke}}y, {}^{S_{yoke}}z) \tag{19}$$

where $Q$ is diagonal matrix of $Q_i$, $I_{motor}$ is column matrix of $I_i$.

Aiming to obtain the coil's position in $Q$, a measuring model was established. As illustrated in Figure 2, the coil was simplified to a particle denoted as $P$. This problem was attributed to getting the position of $P$ described in $S_{yoke}$. The coil's position can be derived as

$$\begin{bmatrix} {}^{S_{yoke}}x_m & {}^{S_{yoke}}y_m & {}^{S_{yoke}}z_m & 1 \end{bmatrix}^T = {}^{S_{yoke}}P_m = ({}^{S_B}_{S_{yoke}}T_m)^{-1} \cdot {}^{S_B}_{S_F}T \cdot {}^{S_F}P_m \tag{20}$$

where ${}^{S_F}P_m$ is the $m$-th (m = 1 to 8) coil's position described in $S_F$. ${}^{S_B}_{S_{yoke}}T_m$ is the matrix transferring $S_{yoke}$ to $S_B$. Both ${}^{S_F}P_m$ and ${}^{S_B}_{S_{yoke}}T_m$ are fixed and can be obtained before system deployment. ${}^{S_B}_{S_F}T$ is the time-varying matrix (equation (12)) and updated by the position measuring model (equation (9)).

## 4.2. Optimum actuation allocation

The relation between resultant actuation and an individual actuator's forces is derived in equation (1). How to find an achievable allocation scheme is an optimization problem (Webster and Sousa, 1999; Johansen et al., 2004).

Large power consumption generates heat, which has an adverse impact on precision equipment. Besides, a space laboratory's power is limited, requiring equipment to be as power saving as possible. Therefore, the allocation scheme must fulfill this requirement. We allocated actuators' current based on a "quadratic programming"

(QP) scheme. An energy cost function can be established as

$$C_{cf} = \frac{1}{2} \cdot \boldsymbol{I}_{motor}^T \cdot \boldsymbol{I}_{motor} = \frac{1}{2} \cdot \left(\boldsymbol{Q}^{-1}(y,z) \cdot \boldsymbol{F}_{motor}\right)^T \cdot \underset{8\times 8}{\boldsymbol{I}} \cdot \left(\boldsymbol{Q}^{-1}(y,z) \cdot \boldsymbol{F}_{motor}\right) \tag{21}$$

where $\boldsymbol{I}_{8\times 8}$ is a unit matrix; considering an actuator's nonlinearity, matrix $\boldsymbol{Q}$ is also introduced to ensure accuracy.

The objective of optimum allocation is to find a solution minimizing power consumptions (equation (21)), while following constraints (equation (1)). The mathematical description is formalized as

$$\begin{cases} \min \frac{1}{2} \cdot \boldsymbol{F}_{motor}^T \cdot \boldsymbol{H} \cdot \boldsymbol{F}_{motor} \\ s.t. \boldsymbol{C}_K \cdot \boldsymbol{F}_{motor} = \boldsymbol{F}_T \end{cases} \tag{22}$$

where $\boldsymbol{H} = \left(\boldsymbol{Q}^{-1}(y,z)\right)^T \cdot \boldsymbol{I} \cdot \boldsymbol{Q}^{-1}(y,z)$, $C_K \in R^{6\times 8}$ is row full rank, $\boldsymbol{F}_{motor} \in R^8$ $\boldsymbol{F}_T \in R^6$. The Lagrange function is obtained as

$$L(\boldsymbol{F}_{motor}, \lambda) = \frac{1}{2} \cdot \boldsymbol{F}_{motor}^T \cdot \boldsymbol{H} \cdot \boldsymbol{F}_{motor} - \lambda^T (\boldsymbol{C}_K \cdot \boldsymbol{F}_{motor} - \boldsymbol{F}_T) \tag{23}$$

Let

$$\nabla_{F_{motor}} L(\boldsymbol{F}_{motor}, \lambda) = 0, \quad \nabla_\lambda L(\boldsymbol{F}_{motor}, \lambda) = 0 \tag{24}$$

Then we get

$$\begin{bmatrix} \underset{8\times 8}{\boldsymbol{H}} & -\boldsymbol{C}_K^T \\ \boldsymbol{C}_K & \underset{6\times 6}{0} \end{bmatrix} \begin{bmatrix} \boldsymbol{F}_{motor} \\ \lambda \end{bmatrix} = \begin{bmatrix} \underset{8\times 1}{0} \\ \boldsymbol{F}_T \end{bmatrix} \tag{25}$$

Finally, the result is obtained as

$$\boldsymbol{F}_{motor} = \begin{bmatrix} \underset{8\times 8}{\boldsymbol{I}} & \underset{8\times 6}{0} \end{bmatrix} \begin{bmatrix} \underset{8\times 8}{\boldsymbol{H}} & -\boldsymbol{C}_K^T \\ \boldsymbol{C}_K & \underset{6\times 6}{0} \end{bmatrix}^{-1} \begin{bmatrix} \underset{8\times 1}{0} \\ \boldsymbol{F}_T \end{bmatrix} \tag{26}$$

Another method was adopted as contrast. Minimizing the maximum element in $\boldsymbol{I}_{motor}$, we utilize the fmincon function in MATLAB to allocate the actuation efforts. For the same input ($F_C$ and $\tau_C$) within 1 second, the power consumption of the two methods is illustrated in Figure 8. Furthermore, Figure 8 shows that the QP method always consumes less energy than maximum-minimization method observably.

### 4.3. Controller design and system simulation

A control architecture has been established utilizing Simscape as shown in Figure 7 (b). It includes an actuator nonlinear model (Sub-section 3.1), MVIP's dynamics (Sub-section 3.2), nonlinearity compensation (Sub-section 4.1), optimum actuation allocation (Sub-section 4.2), and controller.

Although the system has coupling between 6-DOF, decoupled single-input and single-output controllers can regulate this system effectively. Figure 9 shows the controller in the $x$-axis. The remaining five are similar to it. Accelerometers are known to be sensitive to drift, so without additional measures, the floater would simply crash into the stator. The relative position measures provided by PSD were introduced to solve this problem. The floater's position was maneuvered below the target vibration isolation frequency.

Accelerometers on the floater were utilized to counteract disturbance stimulated by crew activities and suppress high frequency resonance caused by the cables' mass distribution. A band-pass filter $W_A$ was

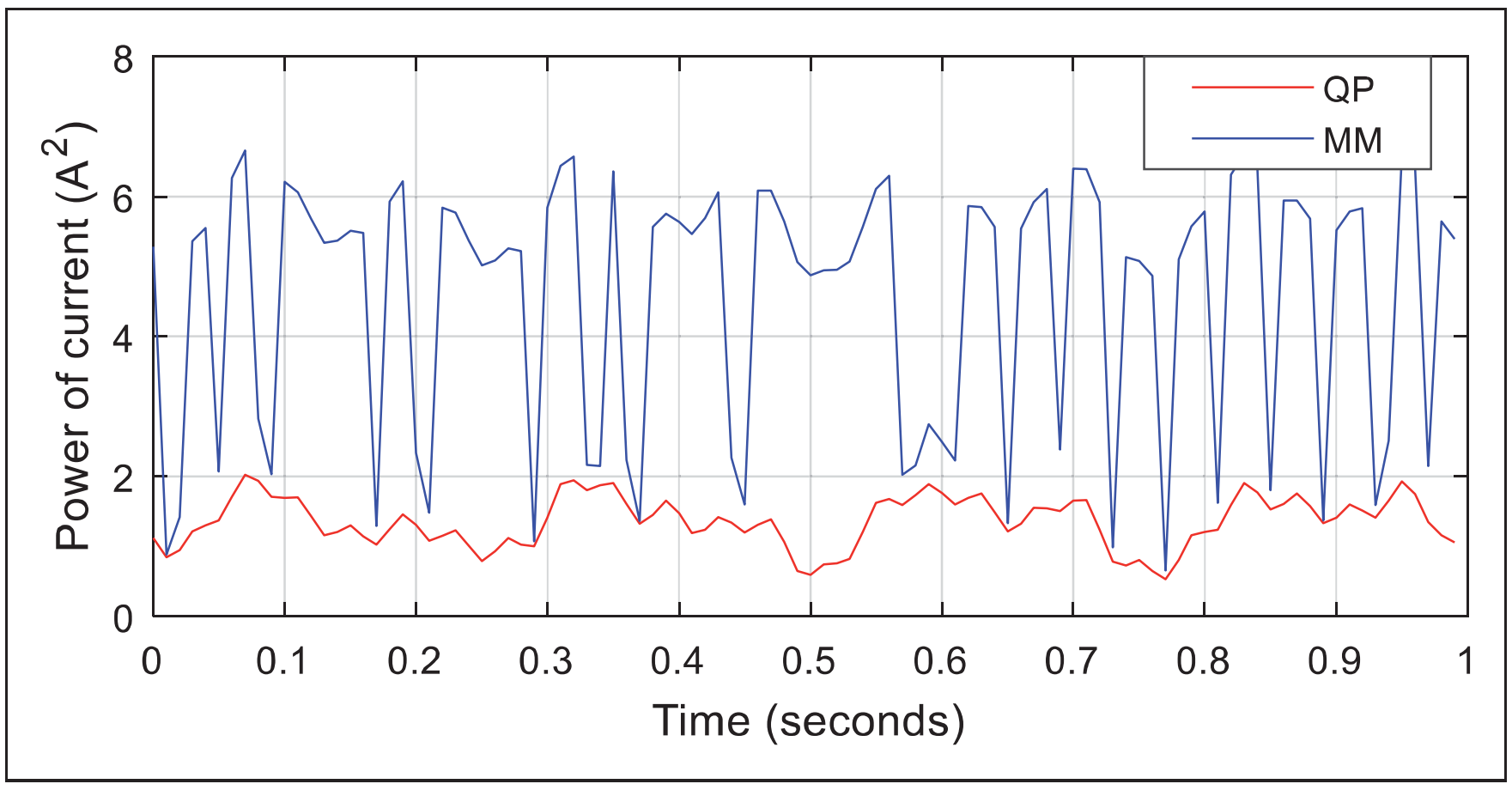

**Figure 8.** Contrast of the two methods' power consumption.

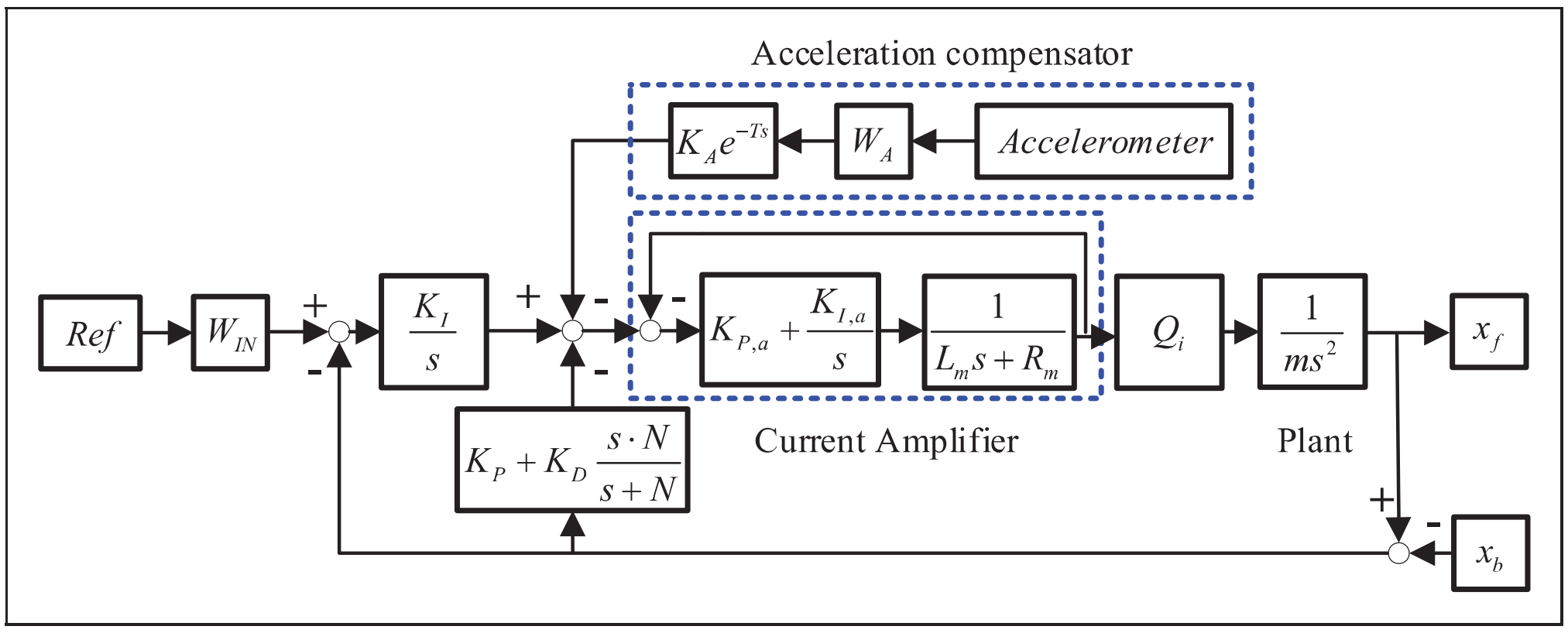

**Figure 9.** I-PD controller with acceleration compensation in *x*-axis.

utilized to eliminate low frequency drift and high frequency noise. It takes the form as

$$W_A = \frac{2\xi\omega_n s}{s^2 + 2\xi\omega_n s + \omega_n^2} \tag{27}$$

where $\omega_n$ denotes the center frequency, $\xi$ denote the pass-band's span, and $s$ the Laplace variable. The filter reaches 1 among the pass-band and at any other frequency $W_A << 1$. The compensation action is only carried out among the pass-band.

The proportional and derivative terms are moved to the feedback loop to avoid the input command's effect. $K_A$ represents acceleration compensation gain and cannot be larger than the floater's mass, in case of forming algebraic-loop and causing instability (Zhu et al., 2006). $e^{-Ts}$ denotes sampling delay, which must be included to represent digital sampling and avoid the algebraic-loop problem. The variables $\ddot{x}_f$ and $\ddot{x}_b$ denote the floater's and stator's acceleration, correspondingly. The transfer function from $\ddot{x}_b(s)$ to $\ddot{x}_f(s)$ is expressed as

$$\frac{\ddot{x}_f(s)}{\ddot{x}_b(s)} = \frac{C(s)}{C(s) + s^2(m + W_A(s)K_A e^{-Ts})} \tag{28}$$

where

$$C(s) = K_P + \frac{K_I}{s} + K_D\frac{s \cdot N}{s + N} \tag{29}$$

The controller's crossover frequency is set at about 0.6 Hz. The proportional term provided needed gain and adjusted the system's natural frequency. For obtaining the desired performance, a derivative action is placed to lift phase margin around crossover frequency. To eliminate the amplification effect of noise caused by the derivative, a low-pass term indicated by $N$ is adopted. Additional criteria are also used to tune this controller depending on requirements. The controller is digitalized by using the zero-order hold method with 2 kHz sampling rate. Figure 10 (a) shows the system's open-loop bode plots in the $x$-axis. The discrepancy of continuous and discrete mainly exists in high frequency and has minor influence on performance. Finally, plant, controller, and complementary sensitivity transfer function are shown in Figure 10 (b). The system's closed-loop cut-off frequency is 1 Hz with a phase margin of 30°. Beyond this frequency, the transmissibility will decay at a ratio of 40 dB/Dec.

### 4.4. Dynamics coupling rectification

The MVIP is controlled based on a decouple scheme. However, installation of the payload will introduce mismatch. After completing the installation of the payload, the CoM and center of gravity (CoG) of the floating body (a combination of floater and payload) will change. Thus, the floater's CoM does not always coincide with its CoG. These will limit the system's performance. Neglecting insignificant terms, such a phenomenon can be modeled as

$$\underset{6\times6}{\boldsymbol{R}_{cp}} \cdot \begin{bmatrix} \boldsymbol{F}_C \\ \tau_C \end{bmatrix} = \begin{bmatrix} \boldsymbol{M} & \underset{3\times3}{0} \\ \underset{3\times3}{0} & \boldsymbol{J}_F \end{bmatrix} \cdot \begin{bmatrix} {}^B\ddot{\boldsymbol{P}}_{FORG} \\ \ddot{\omega}_F \end{bmatrix} \tag{30}$$

where $\boldsymbol{R}_{cp}$ is cross-coupling matrix. Under the decouple scheme, $\boldsymbol{R}_{cp}$ is an identity matrix, which always disagree with fact.

An online rectification scheme is discussed here. We adopted a matrix $\boldsymbol{R}_{rt} = \boldsymbol{R}_{cp}{}^{-1}$ to adjust $\boldsymbol{R}_{cp}$ into identity. Then equation (30) can be rectified as

$$\boldsymbol{R}_{rt} \cdot \boldsymbol{R}_{cp} \cdot \begin{bmatrix} \boldsymbol{F}_C \\ \tau_C \end{bmatrix} = \begin{bmatrix} \boldsymbol{F}_C \\ \tau_C \end{bmatrix} = \begin{bmatrix} \boldsymbol{M} & \underset{3\times3}{0} \\ \underset{3\times3}{0} & \boldsymbol{J}_F \end{bmatrix} \cdot \begin{bmatrix} {}^B\ddot{\boldsymbol{P}}_{FORG} \\ \ddot{\omega}_F \end{bmatrix} \tag{31}$$

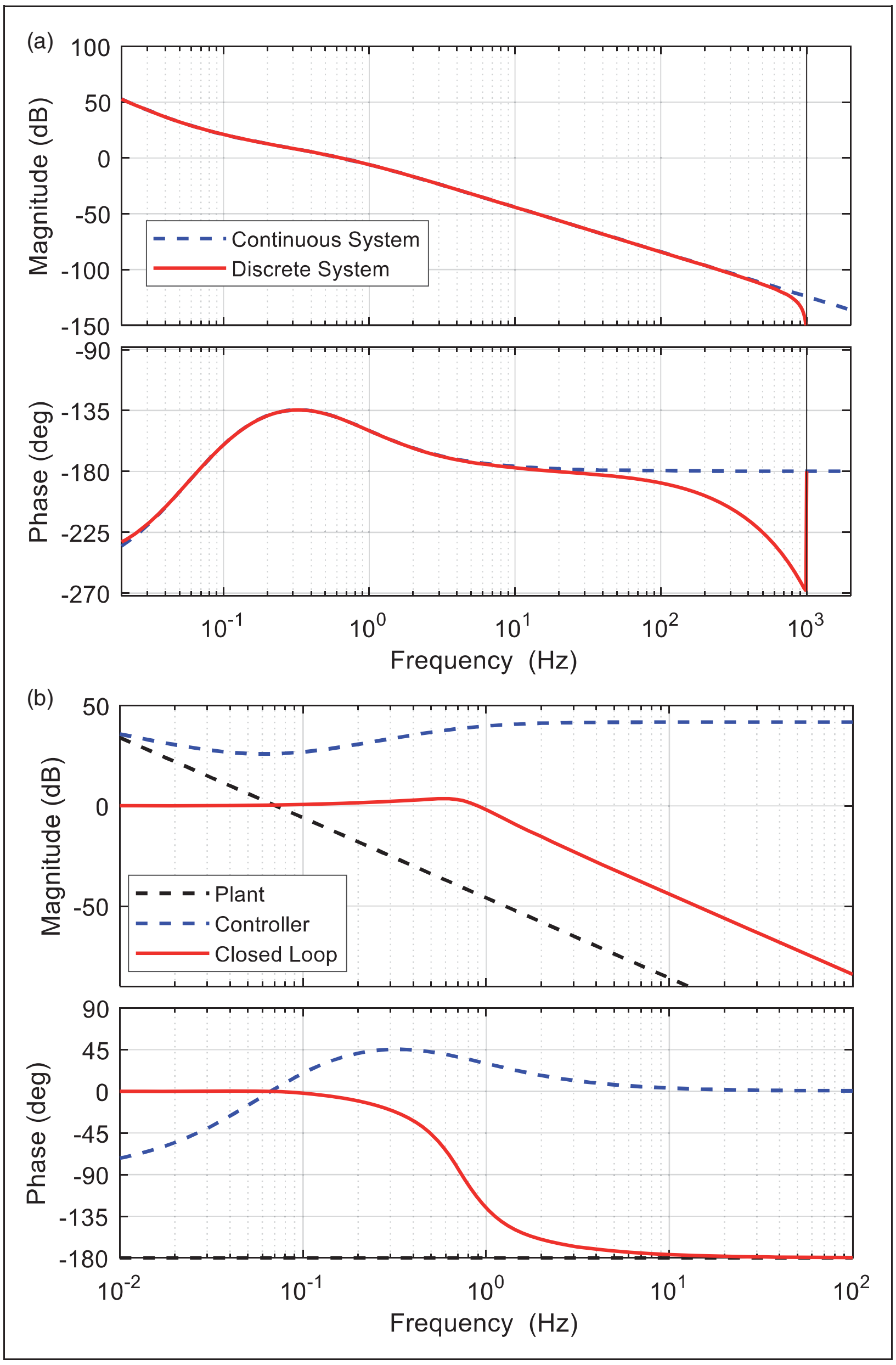

**Figure 10.** Control system simulation analysis: (a) open-loop Bode plots of continuous and discrete system in *x*-axis; and (b) plant, controller, and complementary sensitivity transfer function.

Since the matrix of inertia can be obtained beforehand, how to get $\boldsymbol{R}_{cp}$ is a fitting problem. In order to complete this process automatically, we employed an estimator utilizing RLS in the control-loop. The floater is maneuvered initially by sweep sine instruction within the whole workspace while the recording force commanded and measured acceleration. Data were processed by an estimator online to calculate $\boldsymbol{R}_{cp}$ based on **equation (30)** and then $\boldsymbol{R}_{rt}$ can be calculated and included in the system to rectify system coupling.

**Table 3.** Summary of system performance.

| Specification | Values | Units |
|---|---|---|
| Accelerationnoise | $\leq 1$ | mg |
| Translationrange | $10 \times 10 \times 8$ | mm |
| Translationnoise | 1 | µm |
| Rotationrange | 200 | mrad |
| Rotationnoise | 0.03 | mrad |
| Decayratio | $-40$ | dB/Dec |
| Payload | $\geq 5$ | Kg |

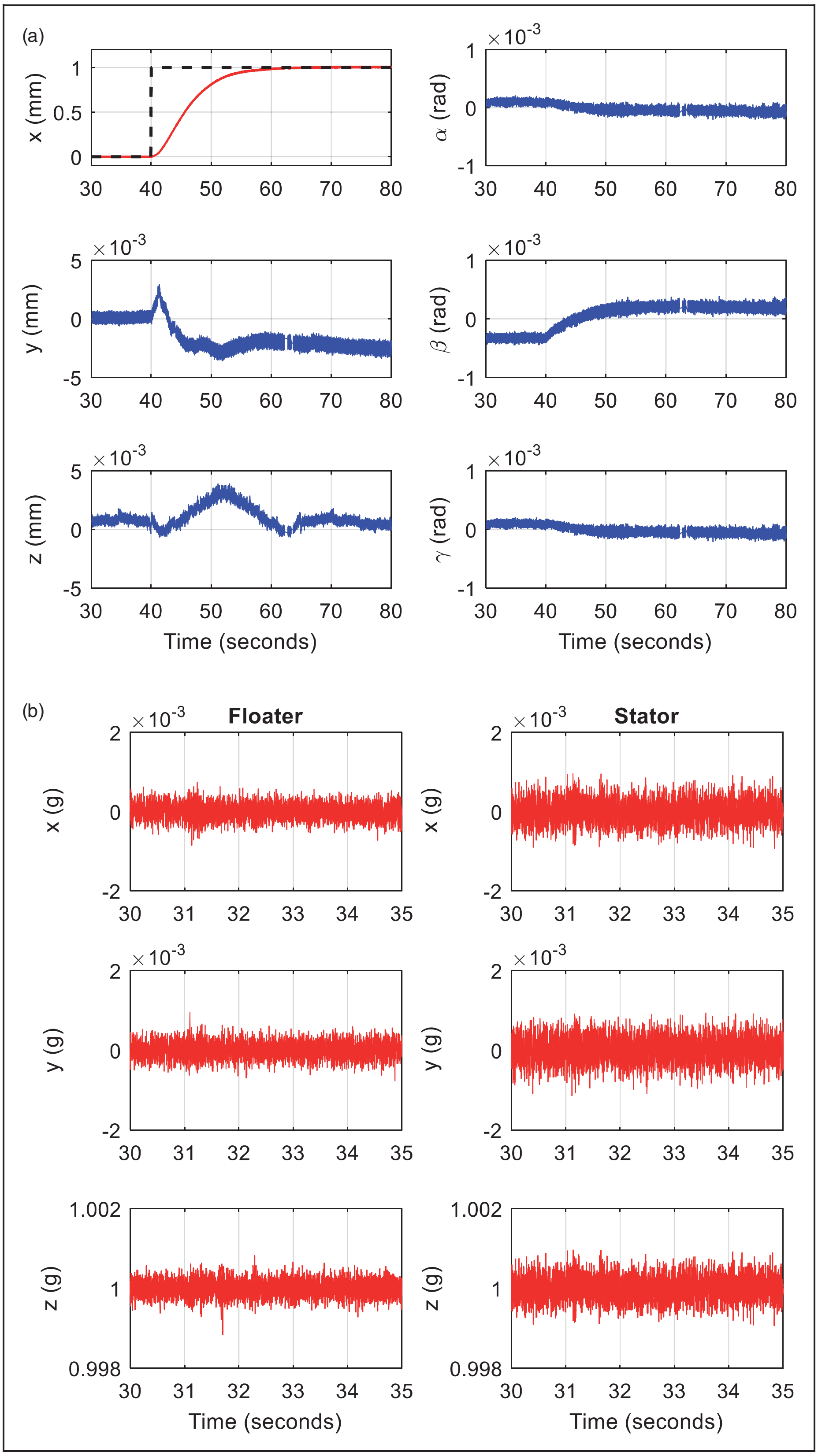

**Figure 11.** (a) Step response in translation along *x*-axis; and (b) steady-state acceleration noise.

## 5. Experimental results

Several experiments were conducted to verify the scheme and system's performance, in view of stabilization, motion range, and vibration isolation. MVIP's performance specifications are summarized in Table 3.

After levitation and stabilization, floater and stator's steady-state acceleration is shown in Figure 11 (b). $z$-axis acceleration is 1 g due to gravity. Peak-to-peak acceleration of all DOF is smaller than 1 mg. Figure 12 shows the frequency spectrum of the floater's acceleration in the $y$-axis. There is a spike at about 55 Hz that can be attributed to power noise. Besides that, the frequency peaks are less than 20 μg in overall frequency.

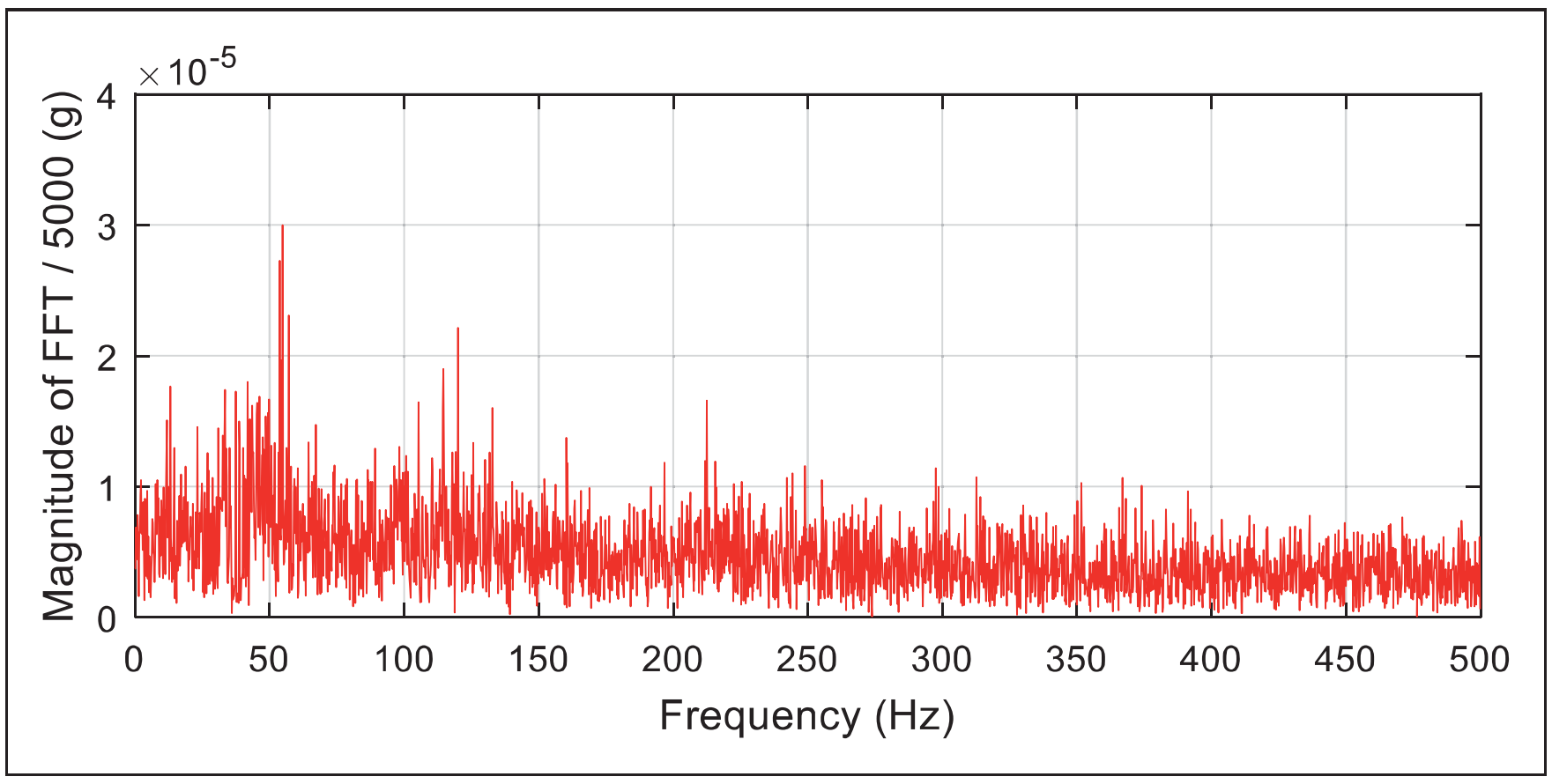

**Figure 12.** Fast Fourier transform (FFT) of *y*-axis acceleration noise on floater.

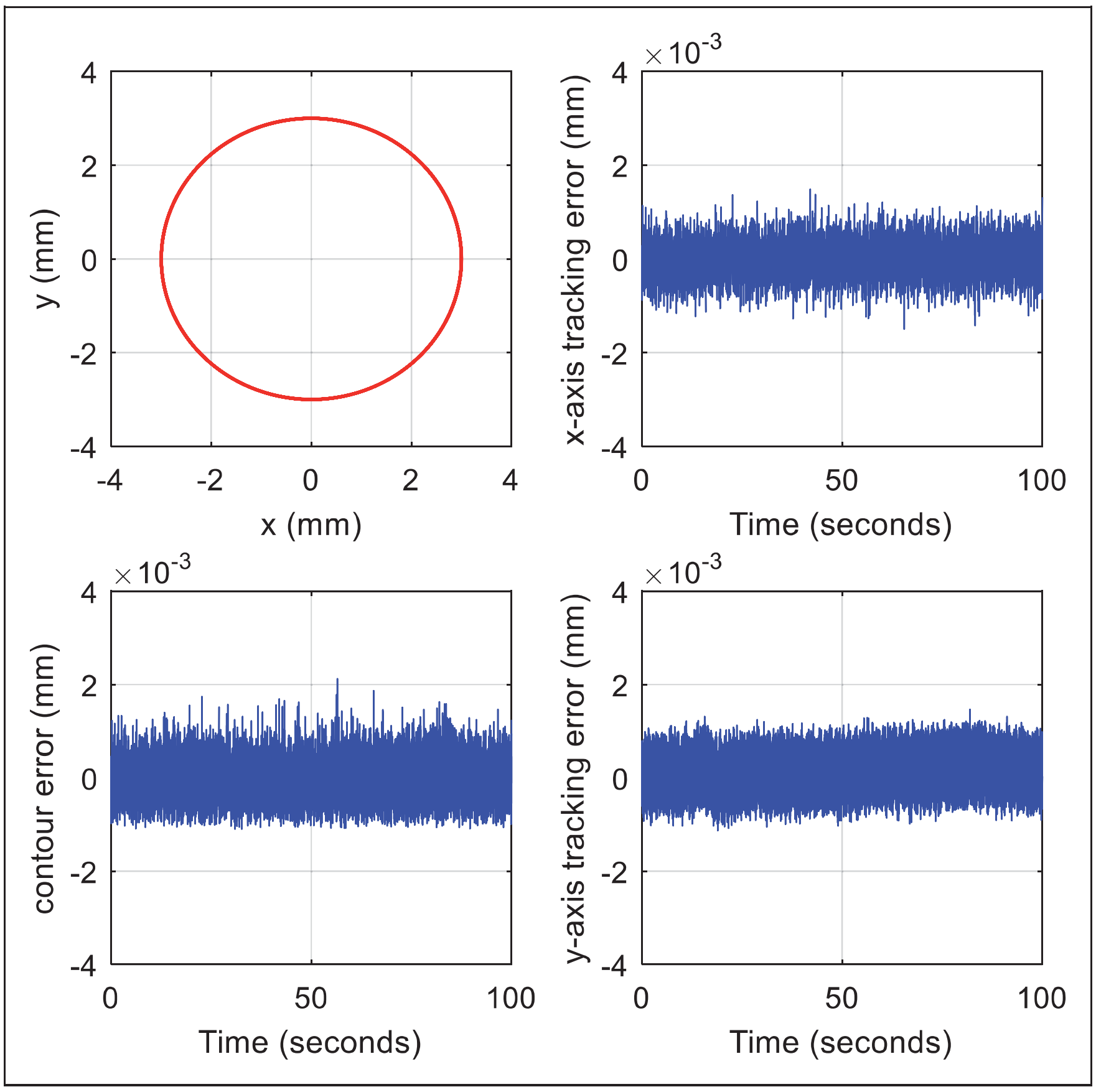

**Figure 13.** Circular contouring control.

Large stoke has beneficial effect in counteracting low frequency vibration. MVIP has a stroke of 10 mm in translation and 20 mrad in rotation. The floater was commanded to contour a circle in the $x$–$y$ plane with a diameter of 6 mm, which relates to the sensor's range. As shown in Figure 13, left top plot depicts the floater's resulting trajectory, right top and right bottom plots depict the tracking error of each axis, and left bottom plot depicts the contour error. The tracking error is less than 2 μm, indicating the accuracy of actuator's mathematical model and the validity of dynamic cross-coupling rectification.

To verify MVIP's vibration suppression ability, the stator must be stimulated by a controllable vibration source. An air floating-based unit was selected in this experiment. This unit consists of optical table, air cushion, and the corresponding holder. As shown in Figure 14, the optical table, covered by glass, provides a reference plane for the air cushion. The air cushion system consists of seven air bearing pads and an air supplying system. The stator and floater are supported by air bearing correspondingly, and an electromagnetic shaker was attached at the unit's margin. Then, MVIP may operate in an artificial nonfriction environment of 3-DOF. Finally, controlled by the shaker, a micro-vibration environment may be obtained.

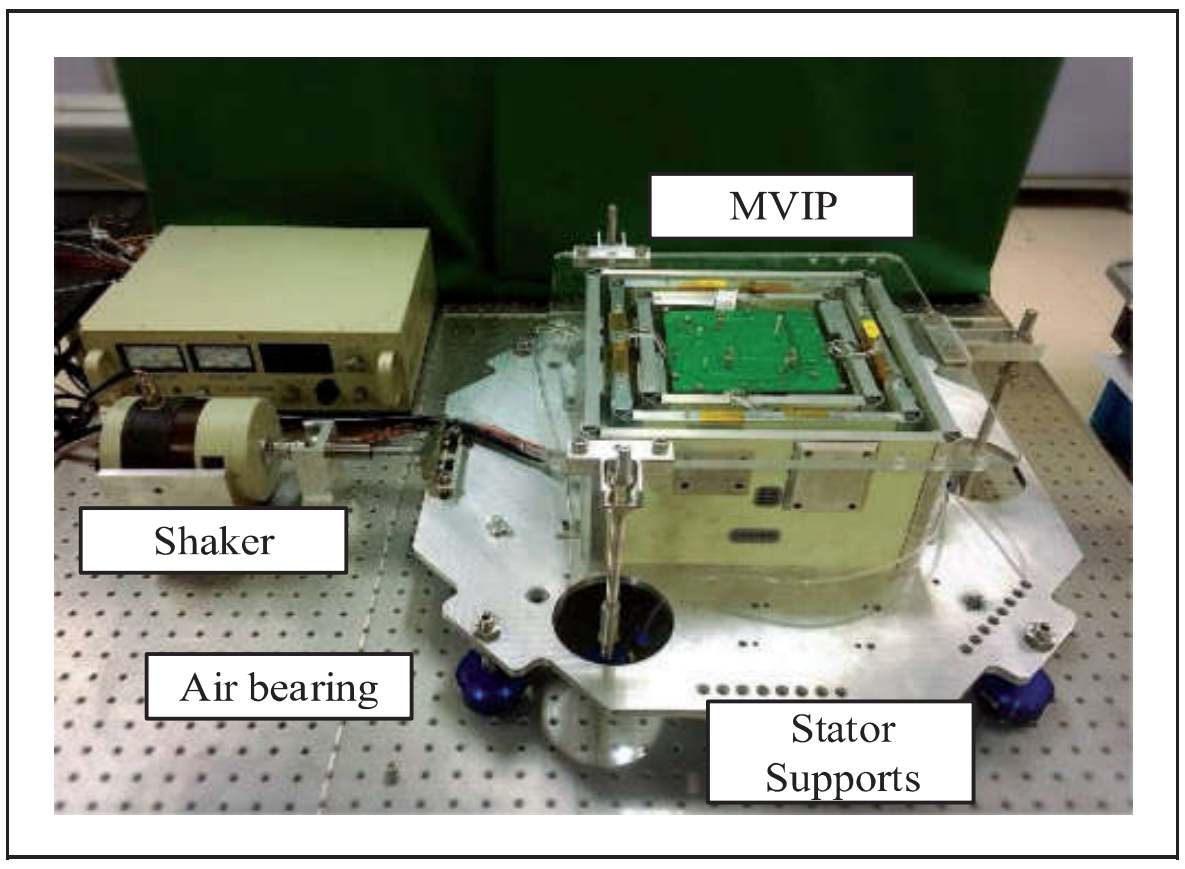

**Figure 14.** Magnetic levitation vibration isolation platform (MVIP) mounted on air cushion system.

Figure 15 shows MVIP's transmissibility along the $x$-axis. In this experiment, the shaker was stimulated by a logarithmic sine sweep signal. Considering the floater's acceleration as output and stator's acceleration as input, both of them were recorded consecutively and compared to obtain the transmissibility. The result shows that MVIP's micro-vibration isolation performance is effective in a wide frequency band with a cut-off frequency of 0.8 Hz. The attenuation rate is about −40 dB/Dec from 1 Hz to 10 Hz. The noise in the high frequency band is mainly caused by the method of experiment and the capability of the micro electro mechanical system type accelerometer. The drifting with lower slope after 20 Hz can be attributed to the unmodeled dynamic behaviors of cables as well as actuators in high frequency, and the dynamic coupling among 6-DOF. The experimental result shows that the designed platform can suppress vibration effectively. Further work will focus on the robust controller design and decoupling method among 6-DOF.

## 6. Conclusion

In this paper, an active vibration isolation platform utilizing maglev theory was designed and made. The concept ensures a large stroke and noncontact operation, which is beneficial to improve the system's vibration isolation performance in a low-frequency band.

There are three main contributions in this paper: (1) a vibration isolation system utilizing maglev theory is proposed; (2) the actuator's mathematical model is established and an optimum allocation scheme is adopted; and (3) the system's nonlinear terms and rectified cross-coupling problem are compensated for.

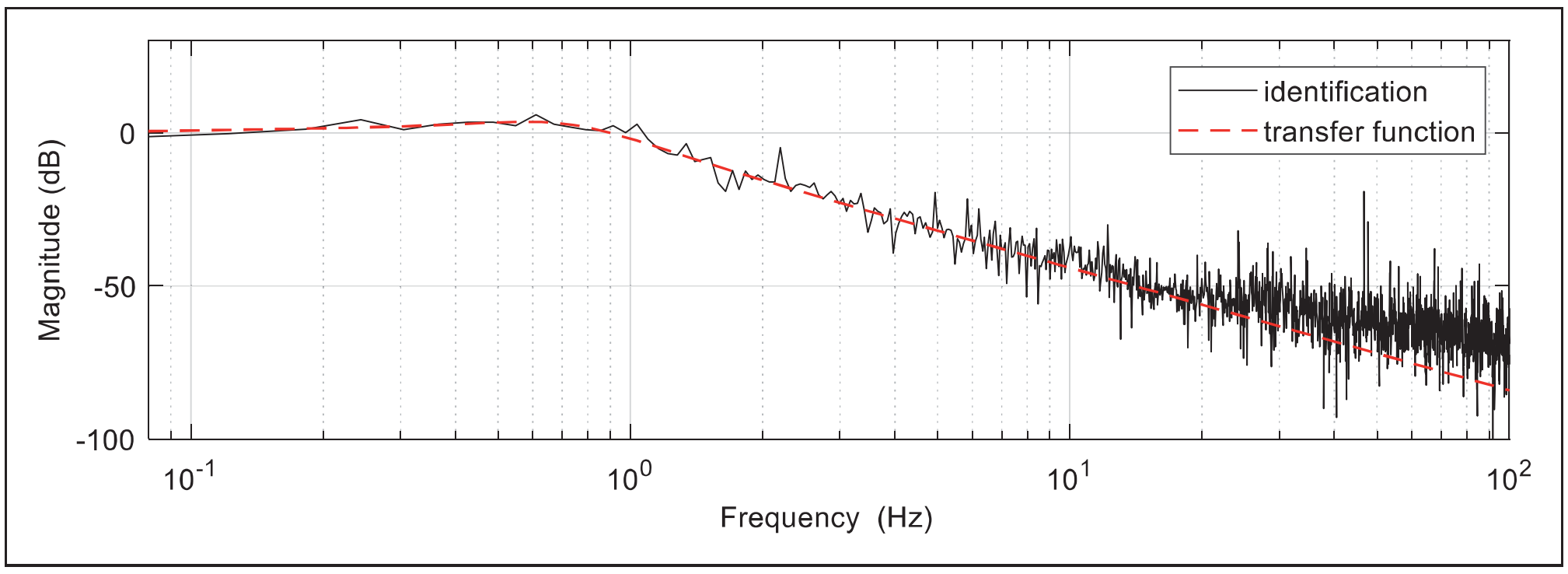

**Figure 15.** Transmissibility of maglev vibration isolation platform from stator to floater along $x$-axis direction.

Based on the established mechatronics model, six independent controllers were adopted and tuned. Several experiments were conducted to illustrate and verify the MVIP's desired performance. A movement range of $10 \times 10 \times 8\,\text{mm}$ and rotations of $200 mrad$ were achieved. Under close-loop control, the statics position noise was less than $1\,\mu\text{m}$ and acceleration noise was less than $20\,\mu\text{g}$. Moreover, a vibration decay ratio with $-40\,\text{dB/Dec}$ between 1 and 10 Hz was also obtained. Future work will focus on reducing cross-coupling between different DOF, improving system performance, and suppressing direct disturbances and ground vibration. With further development, it could become a basis of a feasible, compliant, and adaptable micro-vibration isolation system, deployed in a space laboratory environment.

## Declaration of Conflicting Interests

The author(s) declared no potential conflicts of interest with respect to the research, authorship, and/or publication of this article.

## Funding

The author(s) disclosed receipt of the following financial support for the research, authorship, and/or publication of this article: This study was supported by the National Natural Science Foundation of China (Grant No. 51822502, 51475117), Foundation for Innovative Research Groups of the natural Science Foundation of China (Grant No. 51521003), the Fundamental Research Funds for the Central Universities (Grant No. HIT.BRETIV.201903) and the "111 Project" (Grant No. B07018).

## ORCID iD

Liang Ding http://orcid.org/0000-0002-8351-5178
Honghao Yue http://orcid.org/0000-0003-4334-8632
Yifan Lu http://orcid.org/0000-0002-0661-3372